# Evidence for Half-Quantized Chiral Edge Current in a $C$ = 1/2 Parity Anomaly State

Deyi Zhuo[1,3], Bomin Zhang[1,3], Humian Zhou[2,3], Han Tay[1], Xiaoda Liu[1], Zhiyuan Xi[1], Chui-Zhen Chen[2], and Cui-Zu Chang[1]

[1]Department of Physics, The Pennsylvania State University, University Park, PA 16802, USA

[2]Institute for Advanced Study and School of Physical Science and Technology, Soochow University, Suzhou 215006, China

[3]These authors contributed equally: Deyi Zhuo, Bomin Zhang, and Humian Zhou

Corresponding authors: czchen@suda.edu.cn (C.-Z. Chen); cxc955@psu.edu (C.-Z. Chang).

**Abstract: A single massive Dirac surface band is predicted to exhibit a half-quantized Hall conductance, a hallmark of the $C$ = 1/2 parity anomaly state in quantum field theory. Experimental signatures of the $C$ = 1/2 parity anomaly state have been observed in semi-magnetic topological insulator (TI) bilayers, yet whether it supports a half-quantized chiral edge current remains elusive. Here, we observe a robust half-quantized Hall conductance plateau in a molecular beam epitaxy (MBE)-grown asymmetric magnetic TI trilayer under specific in-plane magnetic field regimes, corresponding to the $C$ = 1/2 parity anomaly state. Within this state, both nonlocal and nonreciprocal transport signals are greatly enhanced, which we identify as direct evidence for a half-quantized chiral edge current localized at the boundary of the top gapped surface. Our numerical simulations demonstrate that this half-quantized chiral edge channel is the essential carrier of the observed half-quantized Hall conductance plateau, analogous to the quantized chiral edge channel in the $C$ = 1 quantum anomalous Hall state. Our results provide experimental evidence for the half-quantized**

**chiral edge transport in a $C = 1/2$ parity anomaly state. This work establishes asymmetric magnetic TI trilayers as a platform for probing single Dirac fermion physics and paves the way to explore a series of exciting phenomena in the $C = 1/2$ parity anomaly state, including the topological magnetoelectric effect and quantized magneto-optical response.**

**Main text:** The search for a single gapless Dirac fermion has been a central pursuit in condensed matter physics[1-3]. In (2+1)-dimensional quantum field theory, a single massless Dirac fermion coupled to an electromagnetic $U(1)$ gauge field breaks parity symmetry upon quantization. This anomalous state is revealed by introducing an infinitesimal mass term, giving rise to a parity anomaly state characterized by a half-integer Chern number and an associated half-quantized Hall conductance $\sigma_{xy}$[4-8]. In condensed matter physics, the concept of the $C=1/2$ parity anomaly state was first proposed in the 1980s[9-11], with the most prominent example being the Haldane model, which predicts the emergence of this state in a monolayer honeycomb carbon lattice under specific valley-tuning conditions[11]. Since then, the parity anomaly state has attracted great interest because it bridges quantum field theory and experimentally accessible condensed matter systems. The parity anomaly state exhibits a distinctive boundary signature associated with a range of exciting physical phenomena, including the topological magnetoelectric(TME) effect[12-16] and quantized magneto-optical response[12,17,18]. Despite decades of theoretical predictions, the experimental realization of the $C=1/2$ parity anomaly state remains elusive, and whether it supports a half-quantized chiral edge current has yet to be explored.

The discovery of three-dimensional(3D) topological insulators(TIs) established a natural platform for exploring parity anomaly physics, as each 3D TI surface hosts a single gapless Dirac fermion[19,20]. When magnetic exchange interactions open a gap on one surface state while leaving the opposite surface gapless, a half-quantized $\sigma_{xy}$ is predicted[21]. Building on this insight, Mogi et al. reported experimental evidence for the $C=1/2$ parity anomaly state in semi-magnetic TI bilayers, where magnetic doping near the top surface introduces a gap while the bottom surface remains gapless[22]. The observed half-quantized $\sigma_{xy}$ at zero magnetic field has been interpreted as an experimental signature of the $C=1/2$ parity anomaly state. However, these signatures are sensitive to an external magnetic field in these semi-magnetic TI bilayers. Establishing a robust half-quantized $\sigma_{xy}$ plateau that reliably hosts the $C=1/2$ parity anomaly state and can be precisely tuned via an external control is therefore essential for exploring single Dirac fermion physics.

In this work, we report convincing evidence for a half-quantized chiral edge current associated with the $C=1/2$ parity anomaly state. By employing molecular beam epitaxy(MBE), we synthesize asymmetric magnetic TI trilayers composed of 3 quintuple layer(QL) V-doped, 6QL undoped, and 3QL Cr-doped $(Bi,Sb)_2Te_3$ layers(Figs. S1 and S2a)[23-25]. Through the application and careful tuning of an in-plane magnetic field $\mu_0H_x$, we achieve a robust half-quantized $\sigma_{xy}$ plateau corresponding to the $C=1/2$ parity anomaly state. Remarkably, we observe strongly enhanced nonlocal and nonreciprocal transport signals near this plateau, providing direct evidence for a half-quantized chiral edge current localized at the boundary of the top gapped surface. Our numerical simulations confirm that this chiral edge channel is the essential carrier of the half-quantized $\sigma_{xy}$ plateau. Our results establish a connection between the $C=1/2$ parity anomaly state and half-quantized chiral edge transport and open new opportunities for investigating single Dirac fermion responses in MBE-grown magnetic TI multilayers.

We first characterize the asymmetric magnetic TI trilayer by performing magneto-transport measurements under an out-of-plane magnetic field $\mu_0H_z$ at the charge-neutral point $V_g=V_g^0=0V$ and $T$=20mK(Fig. 1a). Here $V_g^0$ is defined as the gate voltage $V_g$ at which the longitudinal resistance $\rho_{xx}$ reaches a local minimum in the $C$=1 quantum anomalous Hall(QAH) state at $\mu_0H_z$=0T. The middle 6QL $(Bi,Sb)_2Te_3$ layer serves as a spacer layer that reduces interlayer exchange coupling between the top V- and bottom Cr-doped $(Bi,Sb)_2Te_3$ layers[23,24]. When the magnetizations of the top V- and bottom Cr-doped $(Bi,Sb)_2Te_3$ layers are parallel, the asymmetric magnetic TI trilayer exhibits a well-quantized $C$=1 QAH state. At $\mu_0H_z$=0T, the Hall conductance $\sigma_{xy}(0)$ is ~$1.002e^2/h$, with a corresponding longitudinal conductance $\sigma_{xx}(0)$ of ~$0.017e^2/h$. When the magnetizations of the top V- and bottom Cr-doped $(Bi,Sb)_2Te_3$ layers are antiparallel, the asymmetric magnetic TI trilayer transitions to an axion insulator state[23,24]. This state is characterized by a zero $\sigma_{xy}$ plateau, accompanied by a local minimum in $\sigma_{xx}$. At $\mu_0H_z$=±0.3T, the values of $|\sigma_{xy}|$ and $\sigma_{xx}$ are ~$0.002e^2/h$ and ~$0.003e^2/h$, respectively.

This asymmetric magnetic TI trilayer provides a platform to investigate the half-quantized $\sigma_{xy}$ plateau and the associated $C$=1/2 parity anomaly state under $\mu_0H_x$(Fig. 1b). At $\mu_0H_x$=0T, the magnetizations of the top V- and bottom Cr-doped $(Bi,Sb)_2Te_3$ layers are in parallel alignment. Both surfaces of the middle $(Bi,Sb)_2Te_3$ layer are gapped, each contributing a half-quantized $\sigma_{xy}$. This sample exhibits the $C$=1 QAH state, where the current is carried by a quantized chiral edge channel[26]. As $\mu_0H_x$ increases, the difference in anisotropy fields and the weak interlayer exchange coupling between the top V- and bottom Cr-doped $(Bi,Sb)_2Te_3$ layers allow the magnetization of the bottom Cr-doped $(Bi,Sb)_2Te_3$ layer to tilt in-plane, while the top V-doped $(Bi,Sb)_2Te_3$ layer maintains an out-of-plane magnetization. As a result, the bottom surface state of the middle $(Bi,Sb)_2Te_3$ layer becomes gapless, whereas the top surface state remains gapped[27-32]. Since only one of the paired surface states is gapped, a half-quantized $\sigma_{xy}$ plateau emerges and is identified as a $C=1/2$ parity anomaly state. Given that the side and bottom surface states are gapless, a half-quantized chiral edge channel is predicted to emerge at the boundary of the top gapped surface, according to bulk-boundary correspondence[33,34]. A further increase in $\mu_0H_x$ aligns the magnetizations of both top V- and bottom Cr-doped $(Bi,Sb)_2Te_3$ layers in-plane, rendering both surface states gapless and $\sigma_{xy}$ vanishing, which yields a $C=0$ metallic state(Fig. 1b).

Next, we study how $\mu_0H_x$ affects the QAH and axion insulator states in our asymmetric magnetic TI trilayer, with $\mu_0H_x$ oriented parallel to the applied current direction(Figs. 1c and 1d). To minimize the influence of prior magnetization, $\mu_0H_z$ training is conducted before the $\mu_0H_x$ sweep. As noted above, the $C=1$ QAH state is achieved by sweeping $\mu_0H_z$ from 1.0T to 0T, whereas the $C=0$ axion insulator state is obtained by sweeping $\mu_0H_z$ from 1.0T to -0.3T and then back to 0T. We first initialize the sample into the $C=1$ QAH and $C=0$ axion insulator states. We find that applying $\mu_0H_x$ drives the $C=1$ QAH and $C=0$ axion insulator states into a $C=1/2$ parity anomaly state and ultimately to a $C=0$ metallic state. The $C=1/2$ parity anomaly state emerges when the magnetization of the bottom Cr-doped $(Bi,Sb)_2Te_3$ layer tilts in-plane while the top V-doped $(Bi,Sb)_2Te_3$ layer retains its out-of-plane magnetization(Fig. 1b). In contrast, the $C=0$ metallic state arises when the magnetizations of the top V- and bottom Cr-doped $(Bi,Sb)_2Te_3$ layers are both aligned in-plane(Fig. 1b). The $\mu_0H_x$ values at which the two-step transition occurs are defined as $\mu_0H_{x,t}$ and $\mu_0H_{x,b}$, corresponding to the anisotropy fields of the top V- and bottom Cr-doped $(Bi,Sb)_2Te_3$ layers, respectively.

For $\mu_0H_{x,b}<|\mu_0H_x|<\mu_0H_{x,t}$, a half-quantized $\sigma_{xy}$ plateau is observed with a value of ~$0.504e^2/h$, accompanied by $\sigma_{xx}$~$0.670e^2/h$(Figs. 1c and 1d). The in-plane magnetization of the bottom Cr-doped (Bi,Sb)$_2$Te$_3$ layer can no longer open a gap on the bottom surface of the middle (Bi,Sb)$_2$Te$_3$ layer. As a result, based on parity anomaly-induced boundary excitations, the half-quantized $\sigma_{xy}$ plateau in the $C=\pm 1/2$ parity anomaly states is a local property of the top gapped Dirac surface states[35,36]. The surface-localized half-quantized $\sigma_{xy}$ is independent of the electronic nature of the bottom gapless Dirac surface state and instead emerges from the vanishing Chern number contribution of the bulk states[35]. This interpretation is further supported by our gate-dependent transport measurements(Fig. S4)[25]. The $\sigma_{xy}$ plateau retains half-quantization despite variations in the chemical potential at the bottom gapless surface, provided that the chemical potential is tuned within the magnetic exchange gap of the top gapped surface[22]. Therefore, the half-quantized $\sigma_{xy}$ plateau is a hallmark of the half-quantized surface $\sigma_{xy}$ in magnetic TI films and is closely linked to the bulk TME effect[13,14].

The half-quantized $\sigma_{xy}$ plateau remains stable until the magnetization of the top V-doped (Bi,Sb)$_2$Te$_3$ layer begins to tilt. As $\mu_0 H_x$ increases, $\sigma_{xy}$ gradually decreases and eventually vanishes, while $\sigma_{xx}$ increases to ~1.030$e^2/h$ at $\mu_0 H_x=\pm 6$T. This observation indicates that the magnetization of the top V-doped (Bi,Sb)$_2$Te$_3$ layer tilts in-plane, coinciding with the closure of the magnetic exchange gap in the top surface of the middle (Bi,Sb)$_2$Te$_3$ layer. As a result, the sample exhibits a $C=0$ metallic state(Fig. 1b). The complete two-step transitions from the initial $C=1$ QAH and $C=0$ axion insulator states through the $C=\pm 1/2$ parity anomaly states and finally into $C=0$ metallic state are clearly revealed in the ($\sigma_{xy}$, $\sigma_{xx}$) flow diagrams of the magnetic TI trilayer (Fig. S3)[25]. We note that similar two-step transitions have been observed in magnetic TI pentalayers, i.e., 2QL V-doped (Bi,Sb)$_2$Te$_3$/6QL (Bi,Sb)$_2$Te$_3$/3QL Cr-doped (Bi,Sb)$_2$Te$_3$/6QL (Bi,Sb)$_2$Te$_3$/2QL V-doped (Bi,Sb)$_2$Te$_3$(Fig. S7a)[25]. When the magnetization of the middle Cr-doped (Bi,Sb)$_2$Te$_3$ layer tilts in-plane, a nearly quantized $\sigma_{xy}$ plateau is observed, which can be attributed to the combined contribution of two $C=1/2$ parity anomaly states identified in the asymmetric magnetic TI trilayer[25].

To explore whether the $C=\pm1/2$ parity anomaly states support half-quantized chiral edge transport, we perform nonlocal transport measurements on our asymmetric magnetic TI trilayer(Fig. 2a). The nonlocal resistance is defined as $\rho_{ab,cd}=V_{cd}/I_{ab}$, where $V_{cd}$ is the voltage difference between electrodes $c$ and $d$ and $I_{ab}$ is the current applied from electrodes $a$ to $b$. The initial $C=\pm1$ QAH states are achieved by $\mu_0H_z$ training. For $|\mu_0H_x|<\mu_0H_{x,b}$, both $\rho_{16,34}$ and $\rho_{16,45}$ exhibit a vanishing plateau for $M<0(C=-1)$ and a nonzero plateau for $M>0(C=1)$(Figs. 2b and 2c). The asymmetry of $\rho_{16,34}$ and $\rho_{16,45}$ between $M<0$ and $M>0$ reveals the chirality of edge transport(Fig. 2a). For the $C=-1$ QAH state with $M<0$, the voltage drop across electrodes 2,3,4,5 is negligible because the dissipationless chiral edge channel propagates clockwise(1→2→3→4→5→6), resulting in vanishing $\rho_{16,34}$ and $\rho_{16,45}$. In contrast, for the $C=1$ QAH state with $M>0$, the chiral edge channel propagates anticlockwise from electrodes 1 to 6, so only residual inelastic edge channels contribute to $V_{34}$ and $V_{45}$[37,38], leading to measurable $\rho_{16,34}$ and $\rho_{16,45}$(Figs. 2b and 2c).

For the $C=\pm1/2$ parity anomaly states, we observe a similar asymmetry in $\rho_{16,34}$ and $\rho_{16,45}$ between $M<0$ and $M>0$. For the $C=-1/2$ parity anomaly state with $M<0$, both $\rho_{16,34}$ and $\rho_{16,45}$ are negligible, consistent with transport dominated by a half-quantized chiral edge channel at the boundary of the top gapped surface. However, for the $C=1/2$ parity anomaly state with $M>0$, $\rho_{16,34}$ and $\rho_{16,45}$ are more pronounced than those probed at the $C=-1$ QAH states. This enhancement arises because gauge invariance allows current to inflow through the side gapless surfaces to satisfy the conservation law, forcing the half-quantized chiral edge channel to hybridize with dissipative surface channels on the gapless surfaces. As a result, $\rho_{16,34}$ and $\rho_{16,45}$ remain strongly polarity-dependent due to the concurrence of chiral edge and dissipative gapless surface transport rather than an isolated chiral edge channel. These observations support the existence of the half-quantized chiral edge transport in the $C=\pm1/2$ parity anomaly states.

For the $C=0$ metallic states under $|\mu_0H_x|>3$T, the polarity-dependent differences in $\rho_{16,34}$ and $\rho_{16,45}$ between $M<0$ and $M>0$ disappear due to the absence of edge conduction. The residual nonlocal response in this regime can be attributed to a classical bulk contribution, which can be estimated using the van der Pauw equation $\rho_{NL}/\rho_{xx}=\exp(-\pi L/W)$, where $\rho_{NL}$ is the nonlocal resistance, $L$ is the distance between two voltage electrodes, and $W$ is the channel width[39]. For our Hall bars with $L/W=2$, $\rho_{NL}/\rho_{xx}=\exp(-\pi L/W)=1.87\times10^{-3}$, consistent with the negligible nonlocal resistance observed in the $C=0$ metallic state.

To further demonstrate the existence of the half-quantized chiral edge transport in $C=\pm1/2$ parity anomaly states, we perform direct current(DC) transport measurements to probe chirality-induced nonreciprocity. We apply a DC current $I_{DC}=10$nA and measure $\rho_{xx}$ on the left and right edges $\rho_{xx,L}$ and $\rho_{xx,R}$ for $M<0$ and $M>0$(Figs. 3a to 3c). For the $C=\pm1$ QAH states at $\mu_0H_x=0$T, both $\rho_{xx,L}$ and $\rho_{xx,R}$ vanish and are independent of the magnetization direction, consistent with transport dominated by the quantized chiral edge channel[26]. Figure 3f shows only minor $\rho_{xx}$ differences between $M>0$ and $M<0$ at $\mu_0H_x=0$T, $\Delta\rho_{xx,L}\sim33\Omega$ and $\Delta\rho_{xx,R}\sim-36\Omega$, where $\Delta\rho_{xx,L}=\rho_{xx,L}(M>0)-\rho_{xx,L}(M<0)$ and $\Delta\rho_{xx,R}=\rho_{xx,R}(M>0)-\rho_{xx,R}(M<0)$. This behavior indicates the presence of residual inelastic edge channels in the $C=\pm1$ QAH states, consistent with our nonlocal measurements(Figs. 2b and 2c).

In contrast, the $C=\pm1/2$ parity anomaly states exhibit a pronounced $\rho_{xx}$ sensitivity to both the magnetization direction and electrode configuration. Reversing the magnetization from $M>0$ to $M<0$ produces a substantial decrease in $\rho_{xx,L}$, and a corresponding increase in $\rho_{xx,R}$(Figs. 3d and 3e), violating the Onsager reciprocal relation[40,41]. This nonreciprocal behavior arises from the reversal of the half-quantized chiral edge channel, which causes charge carriers to scatter asymmetrically along a given channel[42,43]. These observations further confirm that the half-quantized chiral edge channel and dissipative surface conduction coexist in the $C=\pm1/2$ parity anomaly states. We note that in the $C=\pm1/2$ parity anomaly states, $|\Delta\rho_{xx,L}|$ and $|\Delta\rho_{xx,R}|$ reach ~765Ω and ~982Ω, respectively(Fig. 3f). The nonreciprocal resistances at the left and right edges have comparable magnitudes but opposite signs, indicating broken inversion symmetry in the DC nonreciprocal transport of the $C=\pm1/2$ parity anomaly states. For the $C=0$ metallic states under $|\mu_0H_x|>3$T, $\rho_{xx,L}$ and $\rho_{xx,R}$ remain identical between $M>0$ and $M<0$, confirming the absence of nonreciprocal transport and ruling out thermal artifacts or higher-order nonlinear contributions in our DC nonreciprocal transport measurements.

To fully understand the half-quantized chiral edge channel of the $C=1/2$ parity anomaly state, we construct a four-band effective 3D TI Hamiltonian[44] with Zeeman splitting $H_\mathrm{M} = \boldsymbol{M}_{t/b}(\mu_0H_x)\cdot\boldsymbol{s}$ on the top and bottom surfaces. The Pauli matrix $\boldsymbol{s}$ operates on electron spin and the magnetization $\boldsymbol{M}_{t/b}$ is controlled by $\mu_0H_x$. The magnetizations remain out-of-plane($\boldsymbol{M}_{t/b} = M_{t/b}\boldsymbol{e}_z$) for $|\mu_0H_x|<\mu_0H_{x,t/b}$, and tilt in-plane($\boldsymbol{M}_{t/b} = M_{t/b}\boldsymbol{e}_x$) for $|\mu_0H_x|>\mu_0H_{x,t/b}$. We calculate the Hall conductance $\sigma_{xy}$, chiral current, and the number of chiral channels $N_c$ of the magnetic TI trilayer under different quantum states[25]. Our numerical simulations show that $\sigma_{xy}$ evolves from $e^2/h$($C=1$ QAH state) or zero(i.e., the $C=0$ axion insulator state), to $e^2/2h$(i.e., the $C=1/2$ parity anomaly state) for $\mu_0H_{x,b}<|\mu_0H_x|<\mu_0H_{x,t}$, and eventually collapses to zero(i.e., the $C=0$ metallic state) for $|\mu_0H_x|>\mu_0H_{x,t}$(Figs. 4a and 4b).

The spatial current distribution reveals that, in both the $C=1$ QAH and $C=1/2$ parity anomaly states, the chiral current is localized at the sample boundaries, confirming the robustness of chiral edge channels(Figs. 4c and 4d). The existence of these channels is a direct consequence of the bulk-boundary correspondence[33,45]. $N_c$ determines the well-defined quantized and half-quantized Hall response, corresponding to the $C=1$ QAH and $C=1/2$ parity anomaly states, respectively(Figs. 4a and 4b). Experimentally, the $C=1/2$ parity anomaly state is metallic, where dephasing effects stabilize the half-quantization[33]. This stands in stark contrast to the traditional quantized Hall effect, which is a fully gapped system. The consistency between theoretical simulations and experimental observations demonstrates that the half-quantized chiral edge channel is the essential carrier of the half-quantized $\sigma_{xy}$ plateau and is directly analogous to the quantized edge channel in the $C=1$ QAH state.

To summarize, we realize a half-quantized $\sigma_{xy}$ plateau and the associated $C=1/2$ parity anomaly states in MBE-grown asymmetric magnetic TI multilayers under in-plane magnetic fields. The observed nonlocal and nonreciprocal transport in the $C=\pm1/2$ parity anomaly states provide convincing evidence for the existence of half-quantized chiral edge channel along the boundary of the top surface. This half-quantized $\sigma_{xy}$ plateau and the half-quantized chiral edge channel arise from the bulk-boundary correspondence and remain robust against parameter variations. Theoretical simulations further confirm that the half-quantized chiral edge channel dominates electrical transport in the $C=\pm1/2$ parity anomaly states. The emergence of the half-quantized $\sigma_{xy}$ plateau offers new insights into the manifestation of the Berry phase in systems enabled by a single

Dirac fermion and paves the way for further exploration of parity anomaly-related phenomena.

**Acknowledgments:** We thank C. X. Liu for helpful discussions**.** This work is primarily supported by the NSF grant(DMR-2241327), including MBE growth and dilution transport measurements. The PPMS measurements are supported by the ONR Award(N000142412133). C.-Z. Chen acknowledges the support from the National Key R&D Program of China Grant(2022YFA1403700), the Natural Science Foundation of Jiangsu Province Grant(BK20230066), and the Jiangsu Shuang Chuang Project(JSSCTD202209). H. Z. is supported by the National Natural Science Foundation of China Grant(123B2058). C. -Z. Chang acknowledges the support from Gordon and Betty Moore Foundation's EPiQS Initiative(GBMF9063 to C. -Z. C.).

**Data availability**: The data that support the findings of this article are openly available [46].

**Figures and figure captions:**

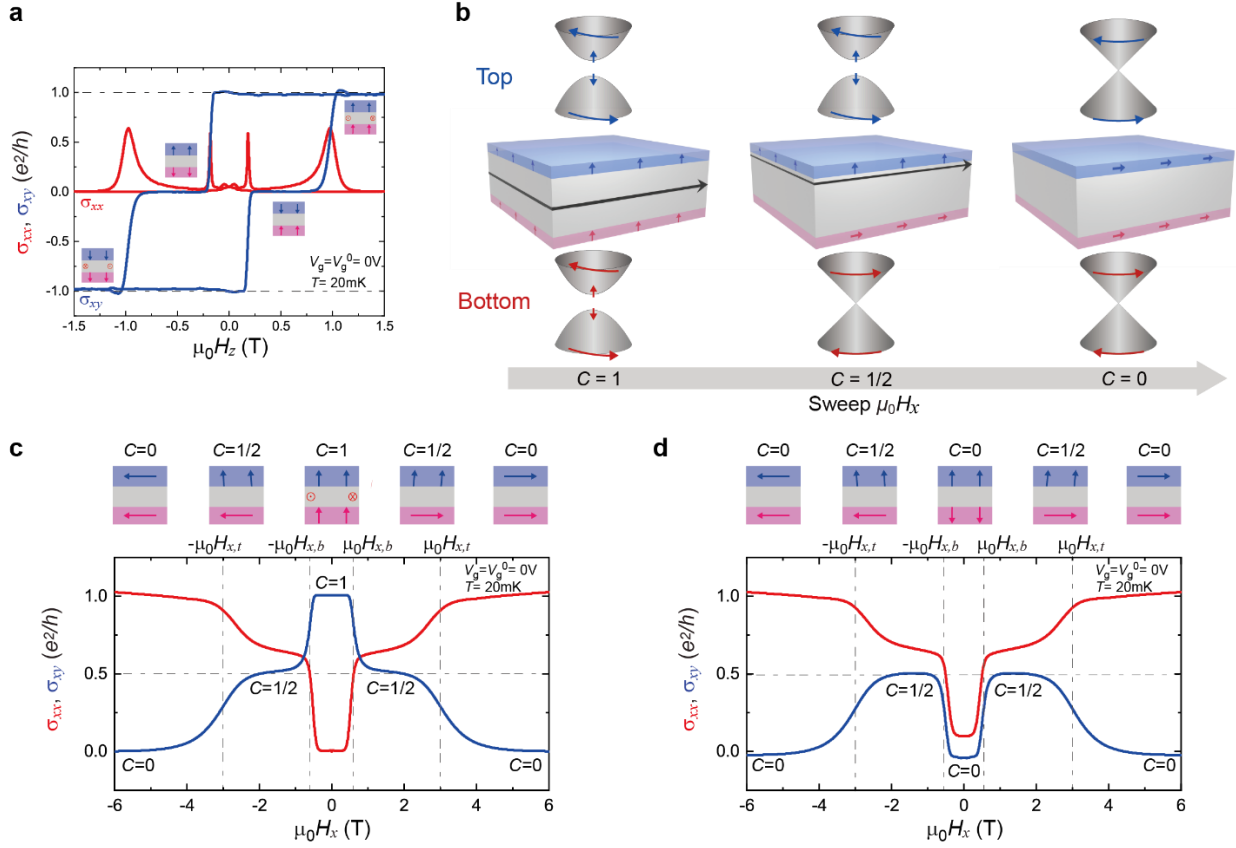

**Fig. 1| $\mu_0H_x$-induced $C=1/2$ parity anomaly state in an asymmetric magnetic TI trilayer. a,** $\mu_0H_z$-dependent $\sigma_{xx}$(red) and $\sigma_{xy}$(blue). **b,** Schematics of the three quantum states with different $C$ as $\mu_0H_x$ is swept. The top blue, middle gray, and bottom red regions represent the 3QL V-doped $(Bi,Sb)_2Te_3$, 6QL undoped $(Bi,Sb)_2Te_3$, and 3QL Cr-doped $(Bi,Sb)_2Te_3$ layers, respectively. **c, d,** $\mu_0H_x$-dependent $\sigma_{xx}$(red) and $\sigma_{xy}$(blue), for systems initialized into the $C=1$ QAH(**c**) and $C=0$ axion insulator(**d**) states. The anisotropy field of the top V-doped $(Bi,Sb)_2Te_3$ layer $\mu_0H_{x,t}$ is ~3.0T, while the anisotropy field of the bottom Cr-doped $(Bi,Sb)_2Te_3$ layer $\mu_0H_{x,b}$ is ~0.5T. All measurements are performed at $V_g=V_g^0=0$V and $T=20$mK

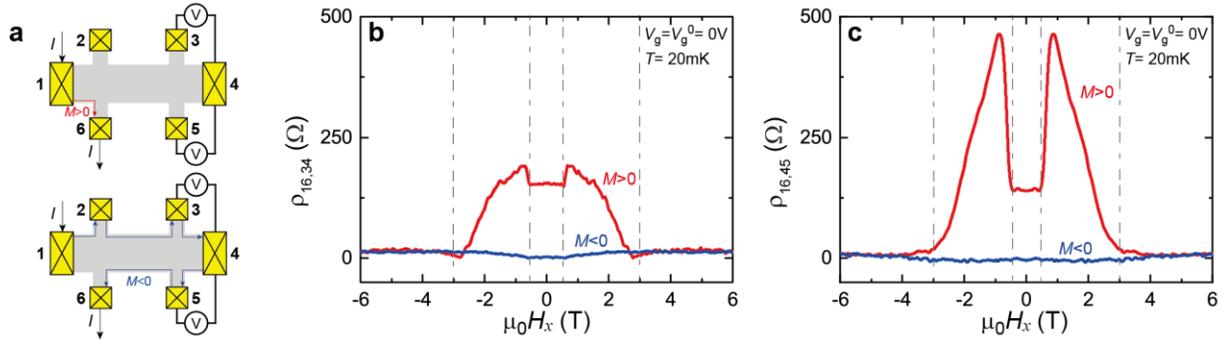

**Fig. 2| Chiral edge transport probed by nonlocal measurements near the half-quantized $\sigma_{xy}$ plateau. a,** Schematics of the nonlocal transport measurements. The red(blue) arrow indicates the direction of the chiral edge channel for $M>0$($M<0$) when the current flows from electrodes 1 to 6. **b, c,** $\mu_0 H_x$-dependent $\rho_{16,34}$(**b**) and $\rho_{16,45}$(**c**) measured at $T=20$mK and $V_g=V_g^0=0$V. The red(blue) curve corresponds to the signal measured for $M>0$($M<0$).

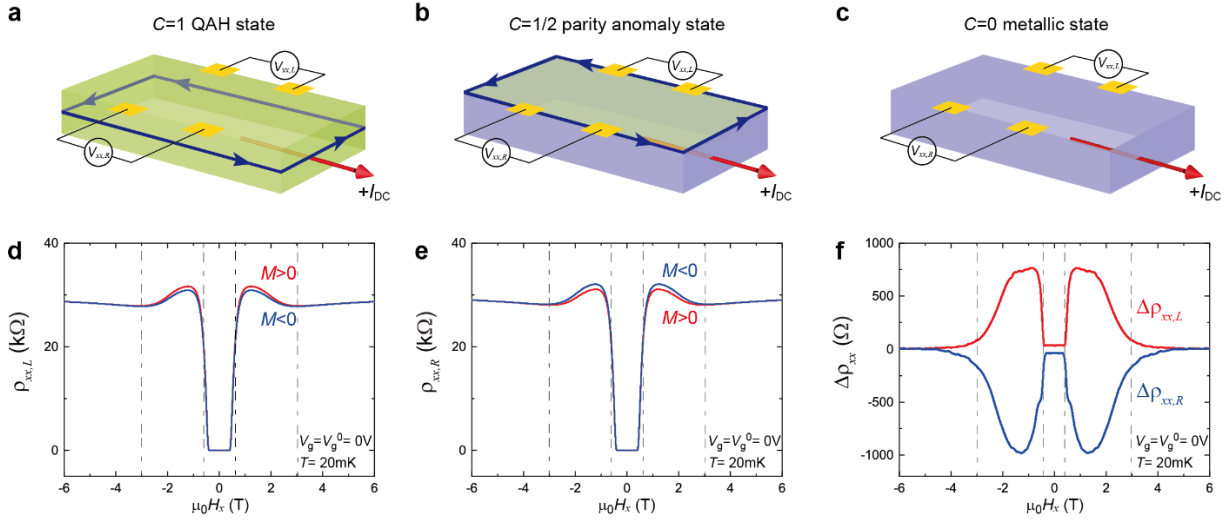

**Fig. 3| Large nonreciprocal transport near the half-quantized $\sigma_{xy}$ plateau. a-c,** Schematics of the $C=1$(**a**), $C=1/2$(**b**) and $C=0$(**c**) quantum states realized under different $\mu_0 H_x$. The dark blue arrows represent the quantized chiral edge channel in the $C=1$ QAH state(**a**) and the half-quantized chiral edge channel in the $C=1/2$ parity anomaly state(**b**). The purple surfaces are the gapless surface states. The red arrow in (**a-c**) is the applied DC current $I_{DC}$. $V_{xx,L}$($V_{xx,R}$) is the longitudinal voltage measured at the left(right) edge. **d, e,** $\mu_0 H_x$-dependent $\rho_{xx,L}$(**d**) and $\rho_{xx,R}$(**e**) measured for $M>0$(red) and $M<0$(blue). $\rho_{xx,L}$($\rho_{xx,R}$) is the longitudinal resistance measured at the left(right) edge. **f,** $\mu_0 H_x$-dependent $\Delta\rho_{xx,L}$ and $\Delta\rho_{xx,R}$, where $\Delta\rho_{xx,L}=\rho_{xx,L}(M>0)-\rho_{xx,L}(M<0)$ and $\Delta\rho_{xx,R}=\rho_{xx,R}(M>0)-\rho_{xx,R}(M<0)$. All measurements are performed with $I_{DC}$=10nA at $T$=20mK and $V_g=V_g^0$=0V.

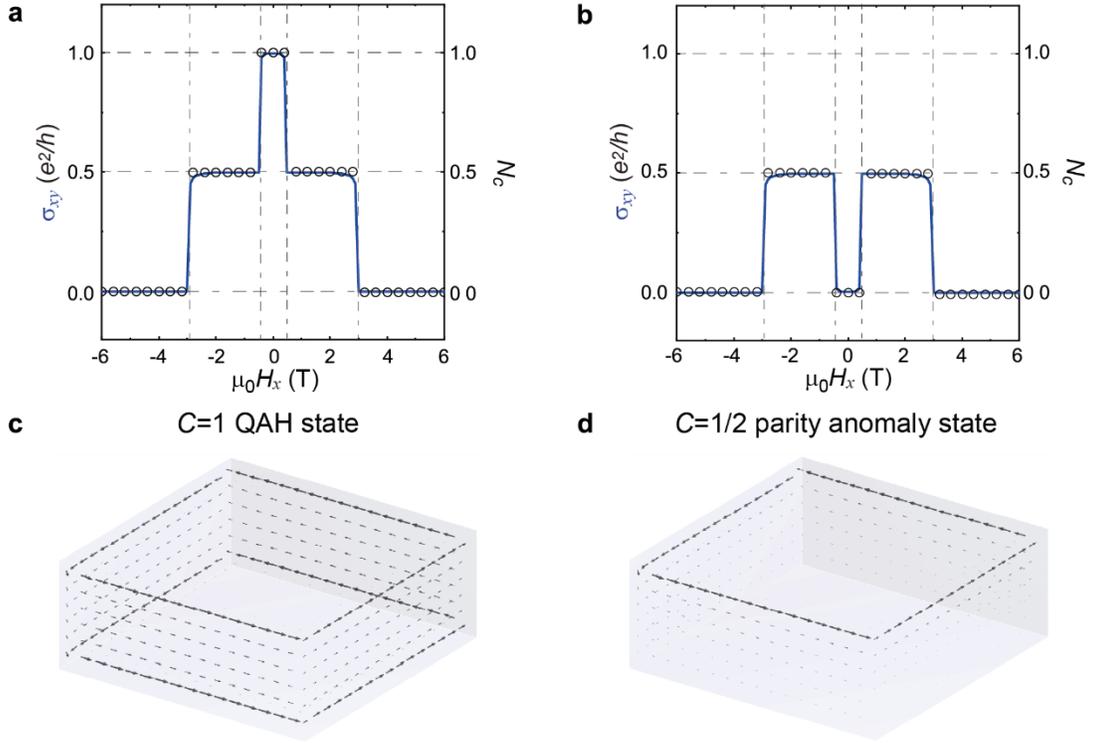

**Fig. 4| Numerical simulations of the half-quantized chiral edge channel in a $C=1/2$ parity anomaly state. a, b,** $\mu_0H_x$-dependent $\sigma_{xy}$ and the number of chiral channels $N_c$, for systems initialized into the $C=1$ QAH(**a**) and $C=0$ axion insulator(**b**) states. **c,** Spatial distribution of the quantized chiral edge current for the $C=1$ QAH state, i.e., the quantum state at $\mu_0H_x=0$T in (**a**). **d,** Spatial distribution of the half-quantized chiral edge current for the $C=1/2$ parity anomaly state, i.e., the quantum state at $\mu_0H_x=1$T in (**a, b**).